\documentclass[aps,prb,preprint,showpacs]{revtex4-1}
\usepackage{graphicx}
\usepackage{amssymb,amsfonts,amsmath}
\usepackage{dcolumn}
\usepackage{color}
\usepackage{subfig}
\usepackage{rotating}
\usepackage{verbatim}

\def\avg#1{\langle #1\rangle }
\begin{document}

\title{Phase Diagram of Boron Carbide With Variable Carbon Composition}
\author{Sanxi Yao}
\author{Qin Gao}
\author{Michael Widom}
\affiliation{Department of Physics, Carnegie Mellon University, Pittsburgh PA  15213}

\date{\today}

\begin{abstract}
Boron carbide exhibits intrinsic substitutional disorder over a broad composition range. The structure consists of 12-atom icosahedra placed at the vertices of a rhombohedral lattice, together with a 3-atom chain along the 3-fold axis. In the high carbon limit, one or two carbons can replace borons on the icosahedra while the chains are primarily of type C-B-C. We fit an interatomic pair interaction model to density functional theory total energies to investigate the substitutional carbon disorder. Monte Carlo simulations with sampling improved by replica exchange and augmented by 2d multiple histogram analysis, predicts three phases. The low temperature, high carbon composition monoclinic $Cm$ ``tilted polar" structure disorders through a pair of phase transitions, first via an Ising-like transition to a ``bipolar" state with space group $C2/m$, then via a first order 3-state Potts-like transition to the experimentally observed ``nonpolar" $R\bar{3}m$ symmetry.
\end{abstract}

\pacs{}

\maketitle

\section{Introduction}
\begin{figure}[ht]
\vskip 0.5cm
\centering\includegraphics[width=0.8\linewidth]{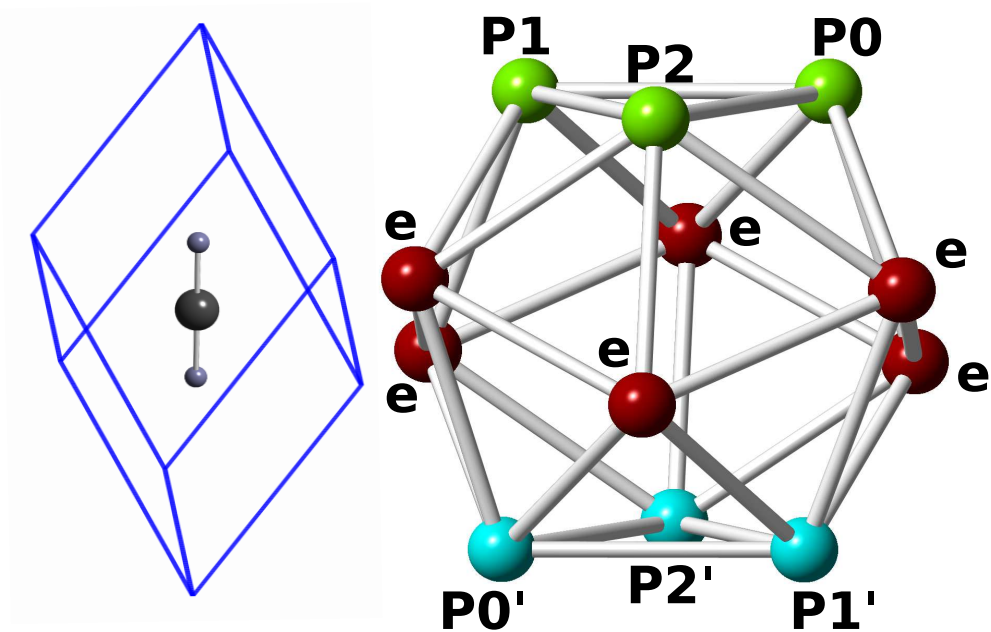}
\caption{Primitive cell of boron carbide showing C-B-C chain at center along the 3-fold axis. The icosahedron (not to scale) occupies the cell vertex. Equatorial sites of the icosahedron are shown in red, labeled ``e'', the north polar sites are shown in green and labeled $p_0$, $p_1$ and $p_2$, while the south polar sites are shown in cyan and labeled as $p_0'$, $p_1'$ and $p_2'$. Equatorial borons bond to chain carbons, while north polar atoms bond to south polar atoms on a neighboring icosahedron.}
\label{fig:structure}
\end{figure}

The experimentally reported phase diagram of boron-carbon~\cite{Schwetz91,Okamoto92} displays three phases: elemental boron and graphite, each in coexistence with boron carbide. The carbon concentration of boron carbide ranges approximately from 9-19.2$\%$ at high temperature.
Crystallographically \cite{GWill76,GWill79,Kwei96,Schmechel00,Widom12}, boron carbide has a 15-atom primitive cell, consisting of an icosahedron and a 3-atom chain, in a rhombohedral lattice with symmetry $R\bar{3}m$. Icosahedra are primarily boron with some carbon substitution, and chains are usually of type C-B-C. The energetically favored electron precise~\cite{Longuet-Higgins55,HZhang2016} structure, with all bonding orbitals occupied, has stochiometry B$_4$C with 20 atomic $\%$ carbon, slightly outside the experimentally observed range. 
The icosahedral carbon preferentially occupies a ``polar" site (see Fig~\ref{fig:structure}).
Icosahedra are connected along edges of the rhombohedral lattice, which pass through the polar sites. For other compositions, the icosahedra can be B$_{12}$, B$_{11}$C$^p$, or even B$_{10}$C$_2^p$ (the bi-polar defect~\cite{Mauri01}), and the chain can be C-B-C, C-B-B, B-B$_2$-B or B-V-B (V means vacancy)~\cite{Yakel,Shirai14,Helmut2016}. Based on density functional theory (DFT) study~\cite{Widom12,Mauri01,Bylander90}, besides the stable monoclinic B$_4$C at the high carbon limit, rhombohedral B$_{13}$C$_2$ is stable with 13.3$\%$ carbon, where every icosahedron is of type B$_{12}$ and every chain is of type C-B-C.

In our previous study~\cite{Yao2015}, we built a pair interaction model at 20$\%$ carbon where each icosahedron contains one polar carbon. The only degree of freedom was the placement of the polar carbon among six polar sites in each icosahedron. We predicted two phases transitions. The low temperature monoclinic ``tilted polar" phase with space group $Cm$ transformed to the ``polar" phase with space group $R3m$ through a first order 3-state Potts-like phase transition and then to the experimentally observed ``nonpolar" rhombohedral phase with space group $R\bar{3}m$ via a continuous Ising-like phase transition.

In this work, we consider variable carbon composition by including B$_{12}$, B$_{11}$C$^p$ and B$_{10}$C$_2^p$ icosahedra. More than two carbons in a single icosahedron is quite rare since it introduces a nearest neighbor C-C bond within the icosahedron that is energetically unfavorable. Similarly, if the global average number of polar carbons exceeds 1, the excess electrons must occupy energetically costly antibonding orbitals~\cite{HZhang2016} or mid-gap states.  Since every chain is C-B-C, we can relate the total carbon composition $x_{\rm C}$ to the mean number of polar carbons per icosahedron $x_p$ as $x_{\rm C}=(2+x_p)/15$.  On energetic grounds we shall constrain $0\le x_p\le 1$.

We build a new pair interaction model with variable carbon composition based on DFT total energies. Simulations on a $2d$ grid of temperature T and chemical difference $\mu\equiv\mu_{\rm C}-\mu_{\rm B}$ use $2d$ replica exchange to better reach equilibrium and overcome large energy barriers between states. We analyze our simulation results with a 2d multiple histogram method~\cite{Swendsen88,Swendsen89}. Similar to our previous study~\cite{Yao2015}, we predict three phases and two phase transitions. However, the intermediate state is now the ``bipolar" state with space group $C2/m$ instead of the ``polar" state with space group $R3m$, signifying the importance of including the bipolar defect~\cite{Mauri01} B$_{10}$C$_2^p$ which was not considered in the old study. We predict the phase diagram at various carbon compositions and characterize the features of these two phase transitions.

\section{Methods}
\subsection{Pair interaction model with variable carbon composition}

\begin{figure}[ht]
\vskip 0.5cm
\centering\includegraphics[width=0.8\linewidth]{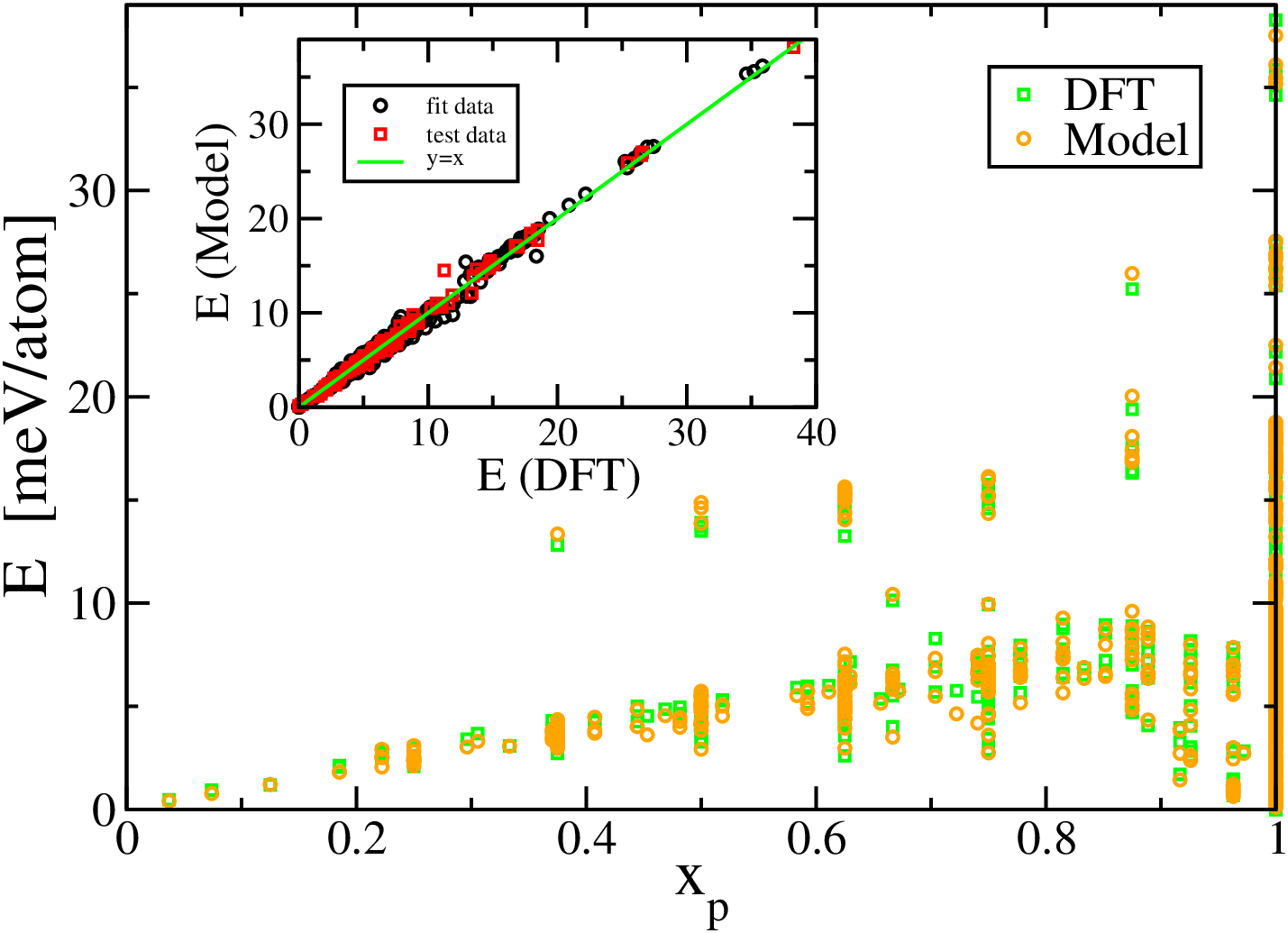}
\caption{Energy $E$ above convex hull calculated from DFT and poly-pair interaction model. Note concave shape with respect to polar carbon concentration $x_p$. Inset shows cross validation of the model with respect to DFT-calculated total energies.}
\label{fig:fit}
\end{figure}

As in our previous study we note the relaxed total energy is a function solely of the initial assignment of carbon atoms to polar sites. In principle we could fit this function with a cluster expansion~\cite{ATAT2002} that is a linear combination of 1-, 2-, and many-body interactions.
Here we use the pair interactions of the cluster expansion, while adding a nonlinear (cubic) function of carbon composition which models the concave shape of energies above convex hull between B$_{13}$C$_2$ and B$_4$C, as seen in Fig.~\ref{fig:fit}. We call this the ``poly-pair interaction model"~\cite{gao2015}. Our model
\begin{equation}
\label{eq:bondmodel}
E(N_1,\dots,N_m)=E_0+\sum_{i=1}^{23} a_k N_k+\beta_0x_p+\beta_1x_p^2+\beta_2x_p^3
\end{equation}
includes 23 types of polar atom pairs ranging from $R_1=1.72$, through the rhombohedral lattice constant $R_9=5.17$ up to $R_{23}=6.58$\AA.

As in our previous study~\cite{Yao2015}, we use the density functional theory-based Vienna ab initio simulation package (VASP)~\cite{Kresse93,Kresse94,Kresse961,Kresse962} to calculate the total energies, of about 600 structures of supercells from 2x2x2 (120 atoms) to 4x4x4 (960 atoms).
Our fitting procedure minimizes the weighted mean-square deviation of model energy from calculated DFT energy, taking an exponential weight related to the energy $\Delta E$ above the convex hull for each structure, so we weight low energy structures more heavily. Five-fold cross validation shows weighted training error around 0.28~meV/atom and weighted test error around 0.31~meV/atom, which corresponds to 15$\times$0.31=4.65~meV/cell, or $k_BT$ per degree of freedom at T=54K. Fig~\ref{fig:fit} illustrates the comparison between DFT and model in one five-fold cross validation.

\subsection{Symmetry and order parameters}
Landau theory allows two symmetry-breaking paths~\cite{Yao2015}, $R\bar{3}m \rightarrow R3m \rightarrow Cm$ and $R\bar{3}m \rightarrow C2/m \rightarrow Cm$, that can be distinguished on the basis of site occupations.
Define $m_0,m_1,m_2,m_{0'},m_{1'},m_{2'}$ as the mean carbon occupancy at polar site $p_0,p_1,p_2,p_0',p_1',p_2'$, respectively. The polar carbon composition
$x_p=\sum_i (m_i+m_{i'})$.
We introduce two order parameters. The longitudinal polarization
\begin{equation}
P_z=m_0+m_1+m_2-m_{0'}-m_{1'}-m_{2'}
\end{equation}
transforms as the one-dimensional irreducible representation $A_{2u}$ of group $D_{3d}$, which breaks inversion symmetry, while preserving rotation and reflection. The pair of functions
\begin{equation}
P_{x}=(m_0+m_{0'})-\frac{1}{2}(m_{1'}+m_{2'}+m_1+m_2), ~~~
P_{y}={\sqrt{3}\over 2}(m_1+m_{1'}-m_2-m_{2'})
\end{equation}
transform as the irrep $E_g$, which breakes rotational symmetry while preserving inversion.
To create a rotationally invariant measure of rotational symmetry breaking we define $P_{xy}=\sqrt{P_{x}^2+P_{y}^2}$.

Both symmetry-breaking paths begin with the fully disordered ``nonpolar" state of highest symmetry $R\bar{3}m$ in which all $m_i$ and $m_{i'}$ are equal ({\em e.g.} $m_i=m_{i'}=1/6$ in the high carbon limit) and $P_z=P_{xy}=0$. Then, in the first path, $m_i\neq m_{i'}$ ({\em e.g.} $m_i=1/3$ and $m_{i'}=0$ in the high carbon limit) gives $P_z>0$ and $P_{xy}=0$, characterizing the ``polar" state $R3m$. Completing the symmetry breaking so that a single polar site is distinguished ({\em e.g.} $m_0=1$ while all others vanish in the high carbon limit), we reach $P_z>0$ and $P_{xz}>0$, characterizing the lowest symmetry ``tilted polar" state, with symmetry $Cm$.
The intermediate state in the second path distinguishes a particular axis $i$ while maintaining the $i\rightarrow i'$ inversion symmetry ({\em e.g.} $m_0=m_{0'}=1/2$ while other $m$'s vanish in the high carbon limit) so that $P_z=0$ and $P_{xy}>0$, characterizing the ``bipolar" phase $C2/m$.
Our previous study~\cite{Yao2015}, where no bi-polar defects were allowed, followed the first path.

\subsection{Monte Carlo simulation and 2d multi-histogram method}
We perform Metropolis Monte Carlo simulations in $L\times L\times L$ supercells of the rhombohedral primitive cell, with $L$ ranging from 3 to 8.  Our simulation includes two types of move: (1) randomly pick a polar site and change the species; (2) randomly interchange a polar carbon site and a polar boron. In view of the high energy cost for occupying antibonding orbitals, we reject moves leading to $x_p>1$.  For each move we calculate the energy change $\Delta E$ and the change in the number of carbon atoms $\Delta N$.  Moves are accepted or rejected according to the Boltzmann factor $\exp(-(\Delta E-\mu \Delta N)/k_BT)$.
Following an equilibration period, we begin recording the total energy $E$ and the occupations $m_i$ ($i=0, 1, 2, 0', 1', 2'$) of the polar sites for each subsequent configuration.

To enhance sampling efficiency we perform $2d$ replica exchange. Consider a set of simulation trajectories; suppose the $i^{th}$ one is at temperature $T_i$ and chemical potential $\mu_i$, with total energy $E_i$ and polar carbon number $N_i$, and similarly for the $j^{th}$ one. The probability of occurence of these two trajectories is proportional to the corresponding Boltzmann factor
\begin{equation}
P_1=P(E_i,N_i;T_i,\mu_i)P(E_j,N_j;T_j,\mu_j)\propto e^{-(E_i-N_i\mu_i)/k_BT_i}e^{-(E_j-N_j\mu_j)/k_BT_j}
\end{equation}
If we swap these two trajectories, so that trajectory $i$ is now at temperature $T_j$ and $\mu_j$, and trajectory $j$ is at temperature $T_i$ and $\mu_i$, then the probability is
\begin{equation}
P_2=P(E_i,N_i;T_j,\mu_j)P(E_j,N_j;T_i,\mu_i)\propto e^{-(E_i-N_i\mu_j)/k_BT_j}e^{-(E_j-N_j\mu_i)/k_BT_i}.
\end{equation}
Detailed balance requires that the acceptance probability to interchange these two trajectories is
\begin{equation}
P=P_2/P_1=e^{\Delta\beta \Delta E-\Delta(\beta \mu)\Delta N}
\end{equation}
where we define $\Delta \beta =1/(k_BT_i)-1/(k_BT_j)$, $\Delta E=E_i-E_j$, $\Delta_N=N_i-N_j$ and $\Delta(\beta \mu)=\mu_i/(k_BT_i)-\mu_j/(k_BT_j)$.

We analyze the Monte Carlo results using $2d$ histograms, similar to the $1d$ analysis in our previous work~\cite{Yao2015}. At a given simulation temperature $T_s$ and chemical potential $\mu_t$, a 2d histogram $H_{T_s,\mu_t}(E,x)$ of configuration energy $E$ and polar carbon composition $x$, can be converted into a density of states~\cite{Swendsen88} $W(E,x)=H_{T_s,\mu_t}(E,x)\exp{((E-Nx\mu_t)/k_BT_s)}$ where $N$ is the total number of atoms. Then we calculate the partition function
\begin{equation}
Z(T,\mu)=\sum_{E,x} W(E,x) e^{-(E-Nx\mu)/k_BT}
\end{equation}
which is accurate over a range of temperatures and chemical potentials close to $T_s$ and $\mu_t$. Free energy F, internal energy U, specific heat $c_v$ and other thermodynamic properties can be obtained directly from Z by differentiation.

\begin{figure}[ht]
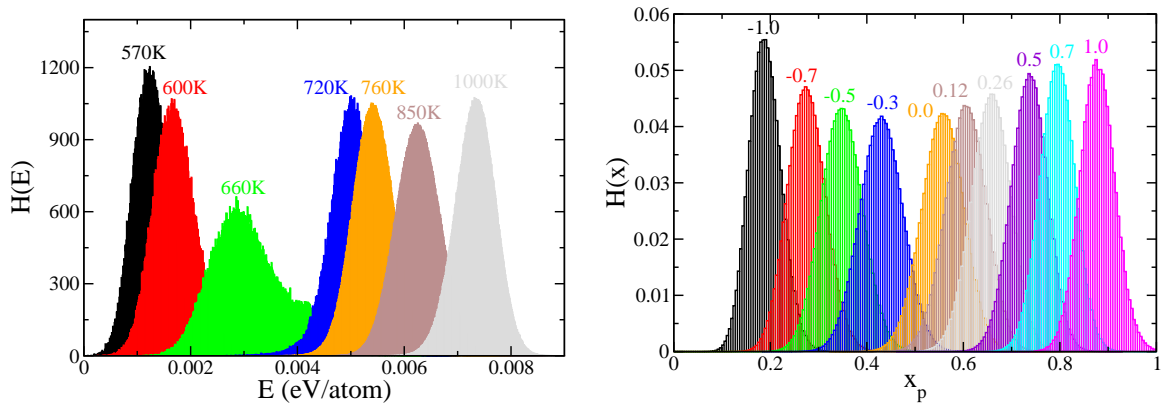

\vskip 0.5cm
\centering
\includegraphics[width=0.45\linewidth]{histE_6x6x6_Mut1-v2.eps}
\hspace{0.1in}
\includegraphics[width=0.45\textwidth]{histnc_6x6x6_T720_Mus-v2.eps}
\caption{Marginal distributions of $H(E,x)$ for 6x6x6 supercell. Left: energy histogram at various Ts fixing $\bar{\mu}=1.0$. Right: polar carbon histogram at various $\bar{\mu}'$s fixing T=720K.}
\label{fig:hist}
\end{figure}

Moreover, by combining histograms taken at temperatures and chemical potentials the density of states can be self-consistently reconstructed~\cite{Swendsen89} so that the free energy becomes accurate over all intervening temperatures and chemical potentails provided the tails of the histograms overlap. Fig~\ref{fig:hist} shows the overlapping marginal distributions of energy histogram for a 6x6x6 supercell at different temperatures fixing $\bar{\mu}=\mu_t/k_BT=1.0$ (left) and marginal distributions of polar carbon composition at different $\bar{\mu}'$s fixing T$_s$=720K (right). The rapid evolution of energy histogram between 660K and 720K indicates a phase transition in this temperature region at this chemical potential.

The 2d density of states $W(E,x_C)$ can be further broken down according to the order parameters $P_z$ and $P_{xy}$, yielding $W(E,x_C,P_z)$ and $W(E,x_C,P_{xy})$, which are joint distribution of energy, chemical potential and the corresponding order parameter. Evaluating powers of these parameters $\langle |P_z| \rangle$, $\langle |P_z|^2 \rangle $, $\langle |P_{xy}| \rangle$ and $\langle |P_{xy}|^2 \rangle$ from the corresponding density of states, we introduce the longitudinal susceptibility $\chi_z$ and in-plane susceptibility $\chi_{xy}$ which are fluctuations of the related order parameters, e.g.
\begin{equation}
\label{eq:chiz}
\chi_z(T,\mu)=N\frac{\avg{|P_z|^2} - \avg{|P_z|}^2}{ k_BT}.
\end{equation}
The susceptibility $\chi_{xy}(T)$ is obtained in a similar way.

\section{Results and discussion}
\subsection{Order parameters}
\begin{sidewaysfigure}
\centering
\includegraphics[trim=0 3cm 0 2cm,clip,width=1.0\linewidth]{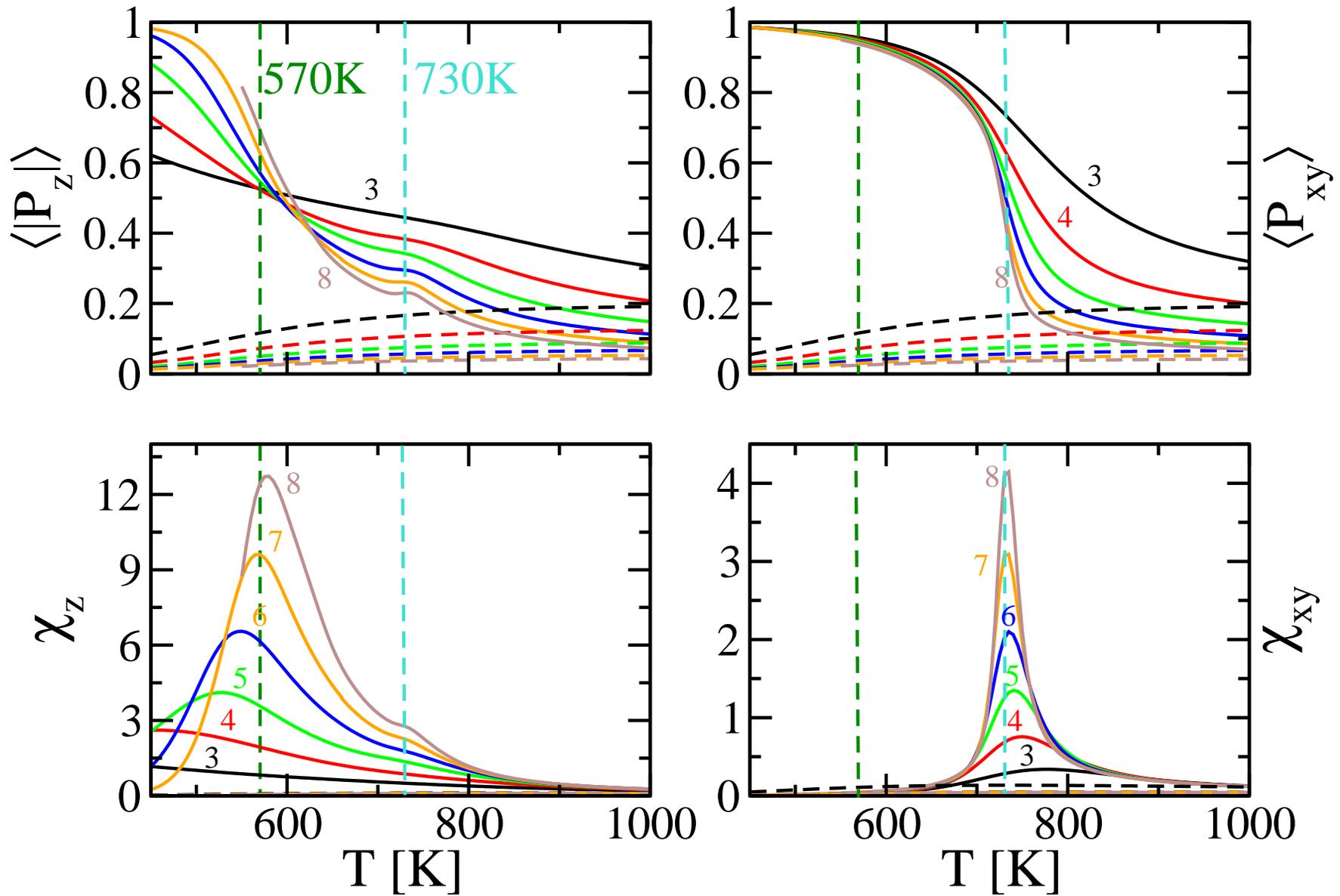}
\caption{Order parameters $\langle |P_z| \rangle$ (upper left), $\langle P_{xy} \rangle$ (upper right) and corresponding susceptibilities $\chi_z$ (lower left), $\chi_{xy}$ (lower right), for supercells from $L$=3 to 8, marked with different colors. Solid curves show order parameters or susceptibilities at high $\mu=0.575$eV, while the broken curves are the corresponding order parameters and susceptibilities at intermediate $\mu=-0.013$eV. }
\label{fig:4pics}
\end{sidewaysfigure}

Plotting order parameters at different temperatures and chemical potentials helps to determine the phases and transitions. An upper limit on chemical potential of $\mu$=0.575eV is determined by comparing the stable structures on the convex hull at zero temperature B$_4$C and graphite. In reality $\mu$ is a function of temperature but we shall neglect this dependence. Fig~\ref{fig:4pics} illustrates order parameters $P_z$ and $P_{xy}$ as a function of temperature, at the high $\mu$ limit (solid) which favors high carbon composition or intermediate $\mu$ (broken) with carbon composition between 0.13 and 0.2. At intermediate $\mu$, both $|P_z|$ and $P_{xy}$ decrease as the system size grows, suggesting that both order parameters vanish at this chemical potential. Then the rhombohedral phase of symmetry $R\bar{3}m$ covers the whole temperature range from 450K to 1000K.

At the high $\mu$ limit where carbon composition approaches 0.2, the average longitudinal polarization $\avg{|P_z|}$ vanishes for $T\gtrsim 570$K but approaches to finite values for $T\lesssim 570$K, while $\avg{P_{xy}}$ decreases with increasing supercell size for $T\gtrsim 730$K  but approaches finite values for $T\lesssim 730$K.  We judge there are three phases, separated by two phase transitions. At high temperature both $P_z$ and $P_{xy}$ vanish and the phase has symmetry $R\bar{3}m$. Below 730K the in-plane polarization $P_{xy}$ grows so one particular direction of $i=0,1,2$ is favored, however $m_i\approx m_{i'}$ so $P_z$ remains small. The phase becomes bipolar with space group $C2/m$ by losing 3-fold rotation symmetry. As temperature decreases further, passing 570K, the longitudinal polarization $P_z$ grows continuously and the phase becomes monoclinic with space group $Cm$, losing inversion symmetry.

\subsection{Specific heat and susceptibility}
The order parameters are first derivatives of the free energy with respect to applied fields, and the corresponding second derivatives are susceptibilities. Specific heat is the second derivative of free energy with respect to temperature. All are evaluated from Monte Carlo data via the fluctuations of energy or order parameters, such as Eq.~(\ref{eq:chiz}).

\begin{figure}[ht]
\vskip 0.5cm
\centering\includegraphics[trim=0 0 0 2.8,clip,width=0.8\textwidth]{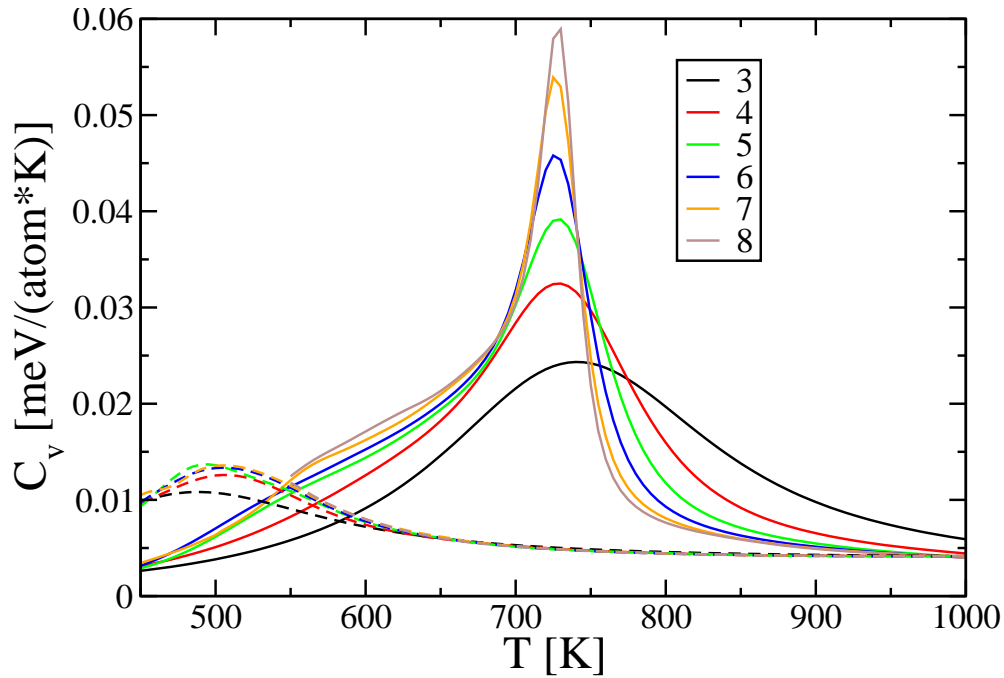}
\caption{Specific heat for L=3 to 8 supercells, at high $\mu$ limit (solid) and intermediate $\mu$ (broken).}
\label{fig:cv}
\end{figure}

Fig~\ref{fig:cv} shows the specific heat for a series of increasing supercell sizes at the high $\mu$ limit and intermediate $\mu$. A strong peak grows with system size around T=730K and another weak peak begins to appear for large system size L=7 around T=570K. These two temperatures coincide to those where the above order parameters change rapidly. Fig.~\ref{fig:4pics} shows the longitudinal and perpendicular (i.e. in-plane) susceptibilities, $\chi_z$ and $\chi_{xy}$ respectively. Both grow with increasing system size. The peak of $\chi_{xy}$ coincides with the strong specific heat peak and big change of $P_{xy}$, while $\chi_{z}$ coincides with the weak specific heat peak and change of $P_z$. 

At intermediate $\mu=-0.013$eV, both specific heat and susceptibility converge to nonzero analytic functions, indicating a single phase with no phase transition. From the order parameters $P_z$ and $P_{xy}$ both of which vanish, we judge this region as a single rhombohedral phase.

\subsection{3-state Potts-like transition at high $\mu$}
As the high temperature transition from $R\bar{3}m$ to $C2/m$ coincides with a breaking of 3-fold rotation symmetry, we expect a three-state Potts-like phase transition which is weakly first order in three dimensions ~\cite{Wu82}. Then the corresponding order parameter $P_{xy}$ should jump discontinuously, and the fluctuations per atom of energy and polarization should grow proportionally to the number of atoms, {\em i.e.} as $L^3$. Our largest supercell size $L=8$ has not yet reached this limit, with divergence around $L^{0.93}$ and $L^{2.6}$ seen for $c_v$ and $\chi_{xy}$ respectively. A similar issue also occurs in our previous work~\cite{Yao2015} and is due to the limited size of $L$.

\begin{figure}[ht]
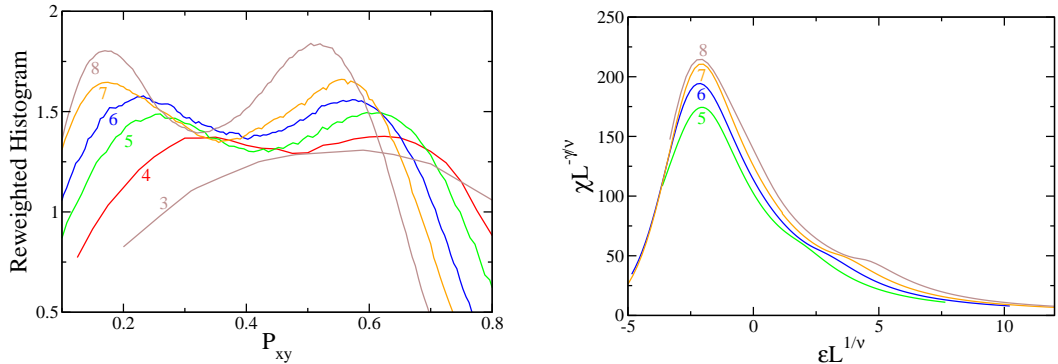

\vskip 0.5cm
\centering
\includegraphics[width=0.4\linewidth]{RWT_with3.eps}
\hspace{0.2in}
\includegraphics[width=0.4\textwidth]{IsingScale.eps}
\caption{Validation of universality classes. (left) Lee-Kosterlitz histograms of $P_{xy}$ for supercells from $L$=3 to 8.; (right) Ising scaling function for $\chi_z$, for supercells from $L$=5 to 8.}
\label{fig:rwt}
\end{figure}

The Lee-Kosterlitz criterion~\cite{Landau05,Lee90} is an alternative method to confirm a first order transition.  Because the two coexisting phases exhibit finite differences in properties such as energy and polarization, probability distributions of such properties should be bimodal, with each peak sharpening as system size grows.  Fig.~\ref{fig:rwt} illustrates this distribution for $P_{xy}$.  This distribution is obtained by marginalizing the joint energy and polarization histogram $H_{T_s}(E,P_{xy})$ over energy, then reweighting with the factor $\exp(E/k_BT_e-E/k_BT_s)$, where the temperature $T_e$ is chosen so as to make the heights of the two peaks equal.  Clearly the distributions of polarization illustrate coexistence of a state with low $P_{xy}< 0.1$ and a state with $P_{xy}\sim 0.5$.  Thus we conclude the transition is first order, as expected for symmetry-breaking of the 3-state Potts type in three dimensions.

\subsection{Ising-like transition at high $\mu$}
As the low temperature transition from $C2/m$ to $Cm$ coincides with a breaking of inversion symmetry, we expect the transition to be in the universality class of the three-dimensional Ising model. Some associated critical exponents are $\alpha=0.110$ (specific heat), $\gamma=1.2372$ (susceptibility) and $\nu=0.6301$ (correlation length)~\cite{Pelissetto02}. Based on finite size scaling theory~\cite{Landau05}, for Ising-like phase transition in three dimensions, the specific heat peak should diverge as $L^{\alpha/\nu}$ ($\alpha/\nu=0.175$), and the susceptibility peak should diverge as $L^{\gamma/\nu}$ ($\gamma/\nu=1.963$).

When plotting the scaled susceptibility $\chi_z/L^{\gamma/\nu}$ as a function of an expanded temperature scale $\epsilon L^{1/\nu}$ with reduced temperature $\epsilon=(T-T_c)/T_c$, Ising universality requires convergence to a common scaling function. As shown in Fig~\ref{fig:rwt} (right), the scaling function curves begin to converge as the system size grows, although with $T_c=627$K, which is higher than the temperature of the peaks of susceptilibity $\chi_z$, implying the systems under study are still too small to show well converged behavior of Ising-like phase transition.

\subsection{Phase diagram in $x_C-T$ plane}
Our simulation on a $2d$ grid of $(\mu,T)$, together with $2d$ multihistogram analysis, allows us to predict the phase diagram in the $x_{\rm C}-T$ plane.
At the high $\mu$ limit it displays three phases and two phases transitions while at the intermediate $\mu$ it displays a single rhombohedral phase.
We do not consider the low $\mu$ case because our model excludes chain variants like C-B-B or B-V-B which are important at low $\mu$, allowing the low $\mu$ rhombohedral phase boundary of rhombohedral phase to extend below $x_{\rm C}=2/15=0.133$. Thus we predict the phase diagram only for $0.133\le x_{\rm C}\le 0.2$.

\begin{figure}[ht]
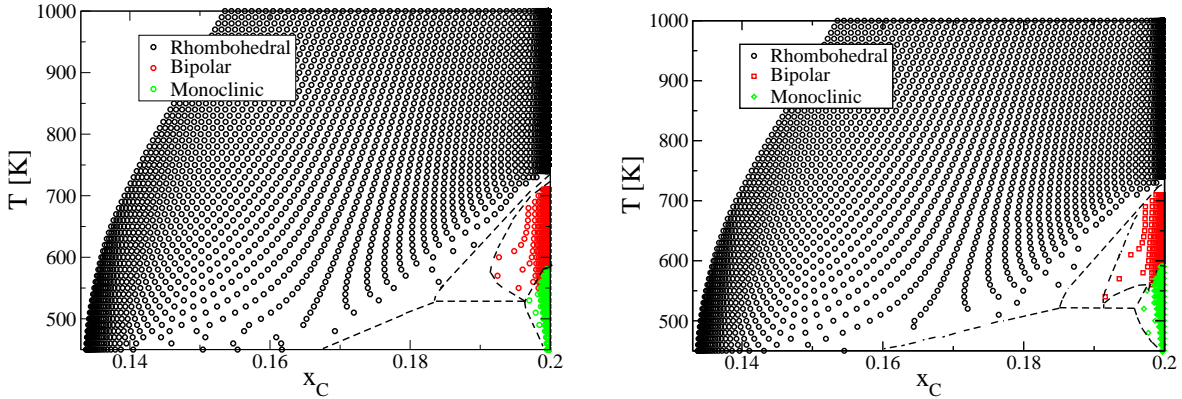

\vskip 0.5cm
\centering
\includegraphics[width=0.45\linewidth]{phases_6_v6-withBoundary.eps}
\hspace{0.2in}
\includegraphics[width=0.45\textwidth]{phases_7_v6-withBoundary.eps}
\caption{Phase diagram prediction from supercell $L$=6 (left) and 7 (right). Thresholds for phases: rhombohedral ($P_z<$0.5 and $P_{xy}<$0.4), bipolar ($P_z<$0.5 and $P_{xy}>$0.6), and monoclinic ($P_z>$0.5 and $P_{xy}>$0.6). Dash curves indicate our qualitatively suggested phase boundaries.}
\label{fig:phase}
\end{figure}

Fig~\ref{fig:phase} shows our predicted phase diagrams based on system size $L$=6 (left) and 7 (right). Threshold values of $P_z$ and $P_{xy}$ define phases based on the corresponding order parameters as a function of temperature (Fig~\ref{fig:4pics}) at the high $\mu$ limit. Black circles indicate the rhombohedral phase $R\bar{3}m$, red squares show bipolar phase $C2/m$, while green diamonds show the monoclinic phase $Cm$. Phase coexistence regions separate the rhombohedral phase from bipolar and from monoclinic, indicating first order phase transitions across the phase boundaries. Even though at high $\mu$ the bipolar phase (red) changes to monoclinic (green) via a continuous Ising like phase transition, at lower $\mu$ the phase diagram implies a gap showing a first order phase transtion, indicating the possible occurence of tricritical point~\cite{Blume1971}. Including variable carbon composition in an Ising-like free energy model could serve as a possible explanation for the change from continuous to first order phase transition as carbon composition decreases, similar to the behavior of the compressible ferromagnet~\cite{Imry1974}.

\section{Conclusion}
We construct a poly-pair interaction model for boron carbide by placing zero, one, or two polar carbons in each icosahedral cluster, introducing B$_{12}$, B$_{11}$C$^p$ or B$_{10}$C$_2^p$ icosahedra, respectively, while fixing the chain to be C-B-C. Our model, which includes pairwise interaction and a nonlinear function of composition, fits DFT total energies well and is amendable to computer simulation. With this model we study the phase diagram and phase transitions over a range of carbon composition. We focus on the range from intermediate to high $\mu$, since we neglect chain variants that are important at low $\mu$.

Monte Carlo simulations on a $2d$ grid of temperatures and chemical potentials, together with $2d$ replica exchange to attain better equilibrium sampling, reveal a wide range of the experimentally observed rhombohedral phase $R\bar{3}m$ at intermediate $\mu$. Three phases occur at the high $\mu$ limit, where two phase transitions occur, one of which at higher temperature is 3-state Potts like, first order, breaking three fold rotation symmetry, and one of which at lower temperature is Ising like, continuous, breaking inversion symmetry. The three phases are rhombohedral ($R\bar{3}m$), bipolar ($C2/m$) and monoclinic ($Cm$), as temperature goes from high to low. These low temperature ordered phases have not been observed experimentally, presumably because atomic diffusion is slow at low T, preventing equilibration.

Globally, we predict the phase diagram in the $x_C-T$ plane, showing regions of the three phases: rhombohedral, bipolar and monoclinic (Fig~\ref{fig:phase}). Empty regions representing the coexistence between phases arise from the first order phase transitions.  At the high $\mu$ limit the transition from bipolar to monoclinic phase is continuous and Ising-like. It could be first order at lower chemical potential due to variable carbon composition, thus a tricritical point may exist.

\section{Acknowledgement}
This work was supported by DOE grant DE-SC0014506.

\bibliography{BC_Poly-prb}

\end{document}